\documentclass[12pt,aps,prd,floatfix,nofootinbib,a4paper,nosuperscriptaddress]{revtex4-2}
\pdfoutput=1

\usepackage{amsmath, amssymb, amsfonts, amsthm, latexsym, epsfig, mathrsfs, xcolor, bbm, slashed, braket, thmtools, cancel}

\usepackage[all]{xy}

\usepackage[inline]{enumitem}

\usepackage{setspace}
\usepackage[marginal, multiple]{footmisc}

\usepackage[T1]{fontenc}
\usepackage[utf8]{inputenc}
\usepackage{lmodern}

\usepackage[colorlinks, allcolors=blue!70!black, linktocpage]{hyperref}

\usepackage{standalone}
\usepackage{tikz}
\usetikzlibrary{decorations.pathmorphing,positioning}

\numberwithin{equation}{section}

\usepackage{cleveref}

\usepackage{microtype}

\usepackage{floatrow}

\let\OLDtableofcontents\tableofcontents
\renewcommand\tableofcontents[1]{%
    {\baselineskip 0.5ex %
	\OLDtableofcontents{#1}}%
}

\let\OLDthebibliography\thebibliography
\renewcommand\thebibliography[1]{%
	\setstretch{1.079} 
	\OLDthebibliography{#1}%
	\small %
	\setlength{\itemsep}{0.2\baselineskip} 
}

\let\OLDfootnote\footnote
\renewcommand\footnote[1]{%
	\setlength{\footnotesep}{0.75\baselineskip}%
	{\footnotesize \OLDfootnote{#1}}%
}

\setlist[enumerate]{noitemsep, label=(\arabic*), ref=(\arabic*)}

\renewcommand\thesection{\arabic{section}}
\renewcommand\thesubsection{\arabic{subsection}}

\makeatletter
\def\p@subsection{\thesection.}
\def\p@subsubsection{\thesection.\thesubsection.}
\makeatother 

\theoremstyle{plain}

\theoremstyle{definition}

\declaretheorem[style=remark,qed=$\scriptstyle{\blacksquare}$,numberwithin=section]{remark} 

\creflabelformat{equation}{#2#1#3}

\crefname{section}{sec.}{sec.}
\crefname{appendix}{appendix}{Appendices}
\crefname{figure}{Fig.}{Figs.}
\crefname{table}{Table}{Tables}

\crefname{definition}{Def.}{Defs.}
\crefname{prop}{Prop.}{Props.}
\crefname{lemma}{Lemma}{Lemmas}
\crefname{corollary}{Cor.}{Cors.}
\crefname{thm}{Theorem}{Theorems}
\crefname{remark}{Remark}{Remarks}

\crefname{ass}{Assumptions}{Assumptions}
\crefname{property}{Properties}{Properties}

\newcommand{\be}{\begin{equation}\begin{aligned}}
\newcommand{\ee}{\end{aligned}\end{equation}}

\newcommand{\lb}{\left}
\newcommand{\rb}{\right}

\newcommand{\mc}{\mathcal}

\newcommand{\ms}{\mathscr}
\newcommand{\mf}{\mathfrak}
\newcommand{\bb}{\mathbb}

\newcommand{\sigop}{\boldsymbol{\delta \sigma}}
\newcommand{\hor}{\mathcal{H}^+}

\newcommand{\eqsp}{\, ,\quad} 

\newcommand{\Lie}{\pounds} 
\newcommand{\defn}{\mathrel{\mathop:}=} 

\newcommand{\union}{\cup} 

\newcommand{\op}[1]{\boldsymbol{#1}}
\renewcommand{\1}{\op{1}}

\let\oldint\int
\renewcommand{\int}{\oldint\limits}

\let\oldlim\lim
\renewcommand{\lim}{\oldlim\limits}

\newcommand{\scri}{\ms I}

\newcommand{\Hilb}{\mathscr{H}}

\newcommand{\antiHilb}{%
\hspace{4pt} 
  \vbox{%
    \hrule height 0.5pt
    \kern0.25ex
    \hbox{%
      \kern-0.3em
      \ifmmode\Hilb\else\ensuremath{\Hilb}\fi
      \kern0em
    }
  }
}

\newcommand{\nfrac}[2]{{{}^#1\!\!/\!_#2}}
\newcommand{\half}{\nfrac{1}{2}}

\newcommand{\abs}[1]{\lb\vert\, #1 \,\rb\vert}		

\DeclareMathOperator{\Tr}{\text{Tr}}

\begin{document}

\setstretch{1.2}

\title{Perturbative semiclassical entropy of dynamical black holes}

\author{Avinandan Mondal}
\email{avinandan@alumni.iitm.ac.in}
\affiliation{Raman Research Institute, Sadashivanagar, Bengaluru 560080, India.}

\author{Kartik Prabhu}
\email{kartikprabhu@rri.res.in}
\affiliation{Raman Research Institute, Sadashivanagar, Bengaluru 560080, India.}

\begin{abstract}
We consider perturbative quantum gravity as a quantum field theory of linearized metric perturbation on an asymptotically flat spacetime with a bifurcate Killing horizon. We include the perturbative gravitational constraints into the algebra of observables restricted to the right half of the future horizon of the spacetime. We use the boundary charge, associated to the horizon Killing field, as an auxiliary ``observer'' degree of freedom. The observables ``dressed'' with the additional charge are invariant under the Killing symmetry and generate  a Type-$\text{II}_{\infty}$ von Neumann factor. We compute the von Neumann entropy of the reduced density matrix of a classical-quantum coherent state constructed from the metric perturbations and the ``observer wavefunction''. This von Neumann entropy satisfies an analogue of the first law of thermodynamics. We further show that this entropy is related to Hollands-Wald-Zhang entropy of the (second order) perturbed dynamical black hole through the flux of perturbations through the horizon and future null infinity.
\end{abstract}

\maketitle
\newpage 
\tableofcontents

\section{Introduction}\label{sec:intro}

Interplay between gravity, thermodynamics and quantum information has long been a key area of investigation to get insights about quantum nature of gravity in the contexts of black hole mechanics and holographic correspondence. In the area of black hole thermodynamics, some of the fundamental insights came from the discoveries that black holes should be assigned an entropy \cite{Bekenstein}, obey laws of thermodynamics \cite{Hawk73} and radiate thermally \cite{Hawk76}. Subsequently it was shown that first law of black hole mechanics is a consequence of general diffeomorphism invariance of gravity and the entropy is the integral of the Noether charge of gravity \cite{Wald, Iyer-Wald}. Study of thermodynamics of quantum field theory in black hole spacetimes dates back to the original work on thermal radiation by Hawking \cite{Hawk76}, but in recent years it has attracted a lot of attention with particular focus on quantum information theoretic quantities (like von Neumann entropy and relative entropy). These have been done by employing powerful techniques of Tomita-Takesaki modular theory and some important results in mathematical physics connected to it: like the type of local QFT algebras and crossed product factors \cite{Araki, Tak73, Tak70} as well as the connection between modular flow and QFT in bifurcate Killing horizons spacetimes \cite{BW, Sewell, Verch}. A large amount of recent work has been done in these areas: the emergence of Type-II factors from Type-III algebras in the context of holographic duality \cite{Witten22, CPW}; gravitationally dressed algebra with respect to an observer in de Sitter space \cite{CLPW}; emergence of Type-II factors by imposing gravitational constraints on algebra of subregions in spacetimes with a bifurcate Killing horizon \cite{Gautam} to name a few.

The issue of entropy of dynamical black holes has also been studied widely in the literature \cite{Iyer-Wald, Wall, HI, HWZ}. In general relativity, the Iyer-Wald entropy \cite{Iyer-Wald} of a black hole at some cut $\mathcal{C}$ of the horizon is the Bekenstein-Hawking entropy $A[\mathcal{C}]/4$ where $A[\mathcal{C}]$ is the area of the cut. A Noether charge prescription for dynamical black hole entropy was given by Wall \cite{Wall} and it was shown in \cite{HWZ} that in general relativity, it matches with Bekenstein-Hawking entropy. Treating external metric perturbations as a quantum field, Hollands and Ishibashi \cite{HI} obtained an expression for relative entropy of a perturbed coherent state with respect to the vacuum state of the black hole and showed that an appropriate variation of this quantity is related to the integral of Bondi news tensor (radiation flux) leaking out through null infinity. Recently Hollands, Wald and Zhang (HWZ) \cite{HWZ} have obtained an entropy for dynamical black holes which corrects the Iyer-Wald prescription with a dynamical correction term. These works raise an interesting problem to compute entropy of a perturbed state of a black hole in the context of linearized quantum gravity. Such a computation will give a perturbative quantum gravitational result for the entropy of dynamical black holes. The resultant entropy is expected to be different from the Noether-charge based entropy prescriptions \cite{Iyer-Wald, Wall, HWZ} as they are the integrals of appropriate $2$-forms at cuts of the horizon and hence they are entropies associated with cuts of the dynamical horizon. However, our entropy expression would be an entropy of the global perturbed state in the entire right future horizon. In this paper, we shall compute that entropy using the mathematical machinery of Tomita-Takesaki modular theory and argue that the resultant entropy can be related to HWZ entropy via a flux term and it is thermodynamic in the sense that it obeys a first law.\\

We consider an asymptotically flat spacetime with metric \(g_{ab}\) and a bifurcate Killing horizon $\mc H = \hor \union \mc H^-$ (see \cref{bif}) generated by the Killing field $\xi^a$ with a constant surface gravity $\kappa$ on the horizon. We denote the inverse temperature of the horizon as $\beta = 2\pi/\kappa$. For the purpose of this paper, we shall assume that the submanifold for whom the surface $\hor \cup i_R^+ \cup \mathscr{I}_R^+$ acts as a Cauchy surface, namely the union of $\mc P$ and $\mc R$ wedges (see \cref{bif}) is globally hyperbolic. Furthermore, we shall assume that for free (linear) fields compactly supported in $\mathcal{P} \cup \mc{R}$, the data can be independently specified on $\hor$ and $i_R^+ \cup \mathscr{I}_R^+$. Subsequently, for the majority of the paper, we shall assume that the data is specified only on $\hor$ and the data is zero on $i_R^+ \cup \mathscr{I}_R^+$. We will relax this last assumption much later in \cref{rem:news-flux} where we will take data on $\mathscr I_R^+$ too.

We then consider linearized metric perturbations $\delta g_{ab}$ supported in the right wedge $\mc R$ of the spacetime (see \cref{bif}). Without loss of generality, we impose certain gauge conditions (see \cref{GeometrySec}) on the perturbations so that the free data of the perturbation on the horizon \(\hor\) is given by the perturbed shear $\delta\sigma_{AB} = \frac{1}{2}\partial_V\delta g_{AB}$ and the first order expansion vanishes $\delta\vartheta = 0$. The phase space of perturbations the horizon is naturally equipped with a symplectic structure (see \cref{symplectic1}) induced by the symplectic current of general relativity. The flow generated by the horizon Killing vector field \(\xi^a\) on this phase space gives a Hamiltonian \(F_\xi\) which can be expressed in terms of perturbed shear as:
\be \label{eq:flux-at-intro}
    F_{\xi} = \frac{\kappa}{4\pi}\int_{\mc{H}^+}dV d\Omega_2 V \delta\sigma_{AB} \delta\sigma^{AB}
\ee
This Hamiltonian \(F_\xi\) measures the flux of gravitational perturbations on the horizon. It can be shown that the flux is the integral of an exact 2-form which depends on \emph{second order} perturbations of the metric \cite{BHBB,HWZ}
\be \label{eq:boundary1}
    F_{\xi} =  \int_{\hor} d[\delta^2Q_{\xi} - \xi\cdot \delta\theta(g, \delta g)]
\ee
where $Q_{\xi}$ is the Noether charge associated to diffeomorphism generated by $\xi^a$ and $\theta$ is the symplectic current in general relativity.

If \(\mc C\) is any cut of the Killing horizon, Hollands, Wald and Zhang (HWZ) \cite{HWZ} have shown that \(\int_{\mc C} \delta^2Q_{\xi} - \xi\cdot \delta\theta(g, \delta g) = \delta^2 S_{\rm HWZ}[\mc C]\) with
\be \label{eq:HWZ-defn-at-intro}
    S_{\rm HWZ}[\mc C] \defn \frac{1}{4}A[\mc C] - \frac{1}{4}\int_{\mc C}d\Omega_2~ V\vartheta
\ee
Here $A[\mc C]$ is the area of the cut $\mc C$, $V$ is the affine parameter along $\hor$ and $\vartheta$ is the expansion. In \cite{HWZ}, \(S_{\rm HWZ}[\mc C]\) is identified as the entropy of a dynamical black hole which produces a dynamical correction to usual area term. From \cref{eq:boundary1}, but integrated between any two cuts of the horizon, it follows that the difference between the HWZ entropy \(\delta^2S_{\rm HWZ}\) at two cuts is equal to the energy flux through the horizon between the two cuts, i.e., the HWZ entropy satisfies a ``physical process'' first law \cite{HWZ}. From this ``physical process'' first law, HWZ proposed that \(\delta^2 S_{\rm HWZ}\) could be viewed as the entropy of a dynamical black hole at leading order in perturbation theory (i.e. at second order for vacuum perturbations). \\

The main goal of this paper is to establish that the HWZ entropy (at second order) is related to a suitable entropy of a perturbed quantum state of linearized gravity on $\hor$. We summarize our construction and results next.

On the horizon we quantize the perturbed shear to get a smeared operator
\be
    \sigop(s) = \int_{\hor}dV d\Omega_2~ \sigop_{AB} s^{AB} 
\ee
where \(s^{AB}\) is a test tensor field on the horizon. This quantization is described in \cref{sec:horizon-quant} and we denote the algebra generated by these perturbed shear operators as $\mc A_{\hor}$. We then consider a centred Gaussian Hadamard state with 2-point function given by \cref{State} as the \textit{vacuum state} $\omega_0$. This state $\omega_0$ when restricted to either side of the bifurcation surface $\mathcal{B}$ is a KMS state\footnote{But note that the state $\omega_0$ is not KMS when restricted to future (or past) of an arbitrary cut since Killing flow does not keep the algebra invariant in the future (or past) of any cut except bifurcation surface.} with inverse temperature $\beta = 2\pi/\kappa$. The flux (classically given by \cref{eq:flux-at-intro}) can also be defined as a operator in this algebra as 
\be
    \op F_{\xi} = \frac{\kappa}{4\pi}\int_{\mc{H}^+}dV d\Omega_2 V :\!\sigop_{AB} \sigop^{AB}\!:~
\ee
where the ``normal ordering'' is with respect to the 2-point function appearing in vacuum expectation value in \cref{State}. This operator satisfies the commutation relation:
\be 
    [\sigop(s), \op F_{\xi}] = -i\sigop(\Lie_{\xi}s)
\ee
and thus generates the Killing flow on the algebra of smeared shear operators.

We denote the GNS Hilbert space obtained from $(\mc A_{\hor}, \omega_0)$ as $\Hilb$. Now in the set of bounded linear operators on $\Hilb$, take the strong (or weak) operator topology closure of Weyl operators supported in the right future horizon $\hor_R$, where the latter is defined as $\op{\mc W}(s): = \exp(i\sigop(s))$ with the smearing tensor $s^{AB}$ now supported in $\hor_R$. The resulting von Neumann factor algebra, denoted by $\mf A(\hor_R, \omega_0)$ is a Type-III factor due to Araki \cite{Araki}. For a nice exposition to von Neumann algebras and their type classification from a physicist's point of view, we refer to \cite{Sorcevon}.

Since the algebra $\mf A(\hor_R, \omega_0)$ is type-III, density matrices and von Neumann entropy are not well-defined in this setting.\footnote{But note that relative entropy between two states is perfectly well defined as it can be defined purely using the state vectors and relative modular operators.} However, in \cite{CLPW} it was shown that if we introduce an observer (moving on a timelike trajectory) into the spacetime and suitably couple the ``observer's degrees of freedom'' to the quantum fields using the perturbative gravitational constraints, the combined \emph{crossed product} algebra is Type-II. This construction can also be described in terms of \emph{quantum reference frames} \cite{FJLRW}. This construction was extended to quantized Klein-Gordan scalar field in asymptotically flat black hole spacetimes in \cite{Gautam} where the asymptotic perturbed ``boundary charges'' in gravity play the role of the ``observer''.

We adapt the analysis of \cite{Gautam} to linearized gravity. In this case the linearized gravitational constraint is $F_{\xi} = X - C$ (see \cref{eq:boundary1}) where $X$ and $C$ are the asymptotic values of the HWZ entropy
\be \label{eq:charges-at-intro}
    X \defn \frac{1}{\beta} \delta^2 S_{\rm HWZ}[i^+_R] \eqsp
    C \defn \frac{1}{\beta} \delta^2 S_{\rm HWZ}[i^-_L]
\ee
The ``observer'' will be the \emph{gravitational charge} $X$.

The charge $X$ in \cref{eq:charges-at-intro} is promoted to an operator $\op X$ acting as multiplication operator on an auxiliary ``observer'' Hilbert space $L^2(\mathbb R)$ and is associated with negative of ``observer Hamiltonian''.\footnote{Note that $\op X$ cannot be directly represented as an operator on the QFT Hilbert space as it depends on the second-order metric perturbations.} Now, on the extended Hilbert space of QFT and observer \(\Hilb^{\rm ext} \defn \Hilb \otimes L^2(\mathbb R)\) we consider the subalgebra of observables which are invariant under the total QFT and observer Hamiltonian. Noting that flux operator $\op F_{\xi}$ is the QFT Hamiltonian, we have the total Hamiltonian as the negative of the constraint $\op C$ 
\be
    \op C = \op X - \op F_{\xi}
\ee
which is now an operator in an extended Hilbert space $\Hilb^{\rm ext}$. The algebra of observables which commute with the constraint operator operator $\op C$ is the crossed product of $\mf A(\hor_R, \omega_0)$ with its modular automorphism group. The modular Hamiltonian for the state $|\omega_0\rangle \in \Hilb$ (where $|\omega_0\rangle$ is the GNS state for $\omega_0$) in the algebra $\mathfrak A(\hor_R, \omega_0)$ coincides with the flux operator(see \cref{BWT}). This crossed product algebra, denoted by $\mf A^{\text{ext}}(\hor_R, \omega_0)$ is a Type-II factor due to Takesaki \cite{Tak73} --- as we argue later it is a Type-$\text{II}_{\infty}$ factor in our case. Hence the crossed prdocut algebra has well-defined desity matrices and a renormalized trace which can be used to compute the von Neumann entropy.

Consider a classical metric perturbation on the horizon $\hor$ represented by the data $\delta g_{AB} = h_{AB}$. This perturbation can be represented by an algebraic coherent state $\omega_h$ on $\mf A(\hor_R, \omega_0)$ (see \cref{coherent}). In the GNS Hilbert space $\Hilb$, the vector representation of the state $\omega_h$ is not unique as it is defined algebraically only in the algebra on the right future horizon (and not on the entire future horizon). However it has a unique representation in the natural cone $\mc P^{\#}$ of $(\mf A(\hor_R, \omega_0), |\omega_0\rangle)$ in $\Hilb$ (see \cref{naturalcone,uniquerep}). We take this unique representation $\ket{ \omega_h }$ in the natural cone as the representation of the perturbed quantum state in $\Hilb$.

We then consider an ``observer wavefunction'' $f(X)$ in the $L^2(\mathbb R)$ space and construct a \emph{classical-quantum} state $|\hat{\omega}_h\rangle$ (as defined in \cref{state}) in the extended Hilbert space $\mathscr{H}^{\rm ext}$. The crossed product factor now being Type-II, one has an well-defined notion of a renormalized trace and one can construct a density matrix corresponding to $|\hat{\omega}_h\rangle$ in the algebra $\mathfrak A^{\rm ext}(\hor_R, \omega_0)$. Let $\op\rho_{\hat{\omega}_h}$ be the density matrix of the reduced state restricted to $\hor_R$. For a `slowly varying'' observer wave-function $f$ the von Neumann entropy of this density matrix can be calculated to be:
\begin{equation} \label{Masteratintro1}
    S(\op\rho_{\hat{\omega}_h}) = -S(\omega_h|\omega_0) + \beta\langle\op{X}\rangle_{\hat{\omega}_h} + S(f) + \log\beta
\end{equation}
where $S(f) = -\int_{\mathbb{R}}dX|f(X)|^2\log|f(X)|^2$ and $S(\omega_h|\omega_0)$ is the relative entropy in the Type-III algebra (\cref{eq:rel-entropy-defn}). Now, it has been shown in \cite{HI, GautamInfo} that the relative entropy $S(\omega_h|\omega_0)$ is given by
\be\label{eq:relentropy-at-intro}
    S(\omega_h|\omega_0) = \beta F_{\xi}[\hor_R] = \frac{1}{2}\int_{\hor_R}dV d\Omega_2 V(\delta\sigma_h)^2
\ee
where $\delta\sigma_h = \frac{1}{2}\partial_Vh$ is the perturbed shear at $\hor$ corresponding to $h_{AB}$ and hence $F_{\xi}[\hor_R]$ is the classical radiation flux falling into the black hole through $\hor_R$.

Combining \cref{Masteratintro1,eq:relentropy-at-intro} as the von Neumann entropy takes the form
\be \label{eq:HWZ-comparison-at-intro}
    S(\op\rho_{\hat{\omega}_h}) = \braket{ \op{\delta^2 S}_{\rm HWZ}[\mathcal{C}]}_{\hat{\omega}_h} - \beta F_{\xi}[\hor_{<\mc C}] + S(f) + \log\beta
\ee
where $\op{\delta^2 S}_{\rm HWZ}[\mathcal{C}]$ is the quadratic form corresponding to the (second order) HWZ entropy at the cut $\mc C$ (\cref{eq:HWZ-defn-at-intro}). Further,$\hor_{< \mc C}$ is the part of the horizon $\hor_R$ between the bifurcation surface $\mc B$ and the cut $\mc C$ and
\be
    F_{\xi}[\hor_{<\mc C}] =  \frac{\kappa}{4\pi}\int_{\hor_{< \mc C}}dV d\Omega_2 V(\delta\sigma_h)^2 
\ee
is the flux through $\hor_{< \mc C}$. Thus, the von Neumann entropy of the density matrix \(\op \rho_{\hat\omega_h}\) for the perturbed state is directly related to the HWZ entropy at some arbitrary cut $\mc C$ of the horizon through a flux of gravitational radiation fallen into the black hole till the cut $\mc C$ along with some state-independent factors ($S(f)$ and $\log\beta$).

In \cref{rem:news-flux} we show that when the radiative flux through null infinity of the perturbation is taken into account we also have
\be \label{eq:radiationtakenintoaccount-at-intro}
    S_{\rm vN}(\op \rho_{\hat{\omega}_h}) = \beta \braket{ \op{\delta^2 H}_R^{\rm ADM}}_{\!\Psi} - \beta F_{\xi}[\hor_R] + \beta F_{\xi}[\scri_R^+] + S(f) + \log \beta
\ee
\(\Psi\) represents the tensor product of the crossed product state on the horizon and the radiative state at null infinity, $F_{\xi}[\scri_R^+]$ measures radiation flux through $\scri^+_R$ and \(\op{\delta^2 H}_R^{\rm ADM}\) is the ADM Hamiltonian associated with \(\xi^a\) at spatial infinity which is defined as a quadratic form.

Our analysis can also be extended to include matter fields in perturbative gravity. For massless fields the state of the matter fields contributes additional flux terms through both the horizon and null infinity. For massive fields there is no flux through null infinity, however there will be an additional contribution to \(X\) (\cref{eq:charges-at-intro}) since massive fields can contribute a charge at \(i^+_R\). We have also neglected any low frequency effects, i.e. ``soft radiation'' or memory effect, on both the horizon and null infinity. We expect that our analysis can be generalized to include these cases following \cite{GautamInfo}.\\

The rest of the paper is organized as follows. In \cref{GeometrySec}, we recall the geometry of bifurcate Killing horizons and vacuum metric perturbations over tham. In \cref{sec:phase-space}, we construct the phase space and classical smeared observables of lienarized gravity on the Killing horizon and detail the quantization on the horizon. We also recall some tools of Tomita-Takesaki theory adapted to our setting and constuct the dressed observables and crossed product algebra. In \cref{EntropySec} we compute the von Neumann entropy of a ``classical-quantum'' coherent state on the horizon and establish the relation to (the second variation) of the Hollads-Wald-Zhang entropy. In \cref{appendix:entropy-calc}, we collect the detailed computation of the von Neumann entropy which are straightforward but rather tedious.

We follow the conventions of \cite{Waldbook}, and work in natural units with $G = k_B = c = \hbar = 1$.. We will work with four dimensional spacetimes though all of our analysis generalizes to other higher dimensions without any modification. We use lowercase Latin letters \(a,b,c,\ldots\) for abstract indices for tensors in the spacetime, e.g., \(g_{ab}\) will denote the spacetime metric and \(\delta g_{ab}\) a linearized metric perturbation. We also use uppercase Latin letters \(A,B,C,\ldots\) for tensors orthogonal to the null generators of the horizon, i.e. on the $2$-sphere cuts of the horizon, e.g. \(\sigma_{AB}\) will denote the shear of the cuts of the horizon. Quantum observables will be denoted by boldfaced version of the symbols for their classical counterparts, e.g. \(\sigop_{AB}\) will denote the quantum observable corresponding to the classical perturbed shear \(\delta\sigma_{AB}\). 

\section{Bifurcate Killing horizons, metric perturbations and gauge conditions}
\label{GeometrySec}

In this section we detail the geometry of the horizon in the background spacetime and detail the gauge condition used on the linearized gravitational perturbations.

\begin{figure}
 \centering
 \begin{tikzpicture}[scale=2.5]

\coordinate (iL0) at (-3,0);
\coordinate (iR0) at (3,0);
\coordinate (iLminus) at (-1.5,-1.5);
\coordinate (iRminus) at (1.5,-1.5);
\coordinate (iLplus) at (-1.5,1.5);
\coordinate (iRplus) at (1.5,1.5);
\coordinate (B) at (0,0);

\draw[thick] (-1.5,-1.5) -- (1.5,1.5);
\draw[thick] (-1.5,1.5) -- (1.5,-1.5);

\draw[black,very thick] (iL0) -- (iLplus);
\draw[black,very thick] (iL0) -- (iLminus);
\draw[black,very thick] (iR0) -- (iRplus);
\draw[black,very thick] (iR0) -- (iRminus);

\node[right] at (0,0) {$\mathcal{B}$};

\draw[black, dotted] 
    (iLplus) -- (iRplus);
\draw[black, dotted] 
    (iLminus) -- (iRminus);

\node at (-0.8,0) {$\mathcal{L}$};
\node at (0,0.8) {$\mathcal{F}$};
\node at (0.8,0) {$\mathcal{R}$};
\node at (0,-0.8) {$\mathcal{P}$};

\node at (-0.9,1.1) {$\mathcal{H}^-_L$};
\node at (0.9,1.1) {$\mathcal{H}^+_R$};
\node at (-1.1,-0.9) {$\mathcal{H}^+_L$};
\node at (1.1,-0.9) {$\mathcal{H}^-_R$};

\node[above left] at (-2.2,0.85) {$\mathscr{I}_L^+$};
\node[below left] at (-2.2,-0.85) {$\mathscr{I}_L^-$};
\node[above right] at (2.2,0.85) {$\mathscr{I}_R^+$};
\node[below right] at (2.2,-0.85) {$\mathscr{I}_R^-$};

\fill[black] (iL0) circle (1.5pt);
\fill[black] (iR0) circle (1.5pt);
\fill[black] (iLminus) circle (1pt);
\fill[black] (iRminus) circle (1pt);
\fill[black] (iLplus) circle (1pt);
\fill[black] (iRplus) circle (1pt);
\fill[black] (B) circle (1pt);

\node[left] at (iL0) {$i_L^0$};
\node[right] at (iR0) {$i_R^0$};
\node[below] at (iLminus) {$i_L^-$};
\node[below] at (iRminus) {$i_R^-$};
\node[above] at (iLplus) {$i_L^+$};
\node[above] at (iRplus) {$i_R^+$};

\draw[->, dashed] (-0.7,0.8) .. controls (0,0.2) .. (0.7,0.8);
\draw[->, dashed] (0.8,-0.7) .. controls (0.2,0) .. (0.8,0.7);
\draw[->, dashed] (0.7,-0.8) .. controls (0,-0.2) .. (-0.7,-0.8);
\draw[->, dashed] (-0.8,0.7) .. controls (-0.2,0) .. (-0.8,-0.7);

\end{tikzpicture}
 \caption{An asymptotically flat spacetime diagram depicting a bifurcate Killing horizon. The flows of the Killing field locally around the bifurcation surface are shown by the curved dashed arrows. The Killing horizon is $\mathcal{H} = \mathcal{H}^+ \cup \mathcal{H}^-$ and divides the spacetime into four regions: left wedge ($\mc{L}$), right wedge ($\mc{R}$), past wedge ($\mc{P}$) and future wedge ($\mc{F}$) as shown. Furthermore the Killing horizons are divided into left ($\mathcal{H}_L^+, \mathcal{H}_L^-$) and right ($\mathcal{H}_R^+, \mathcal{H}_R^-$) sections. $i^+$ and $i^-$ are future and past timelike infinities respectively, $\mathscr{I}^+$ and $\mathscr{I}^-$ are future and past null infinities and finally $i^0$ is spacelike infinity with a subscript of $L$ and $R$ on all of them to differentiate between the left and right wedges respectively. The dotted lines connecting the timelike infinities in past and future can be singularity for Schwarzschild or it might not be the end of the spacetime, e.g. for Kerr black hole. Note that we require $\mc{P} \cup \mc{R}$ to be globally hyperbolic.} 
 \label{bif}
\end{figure}
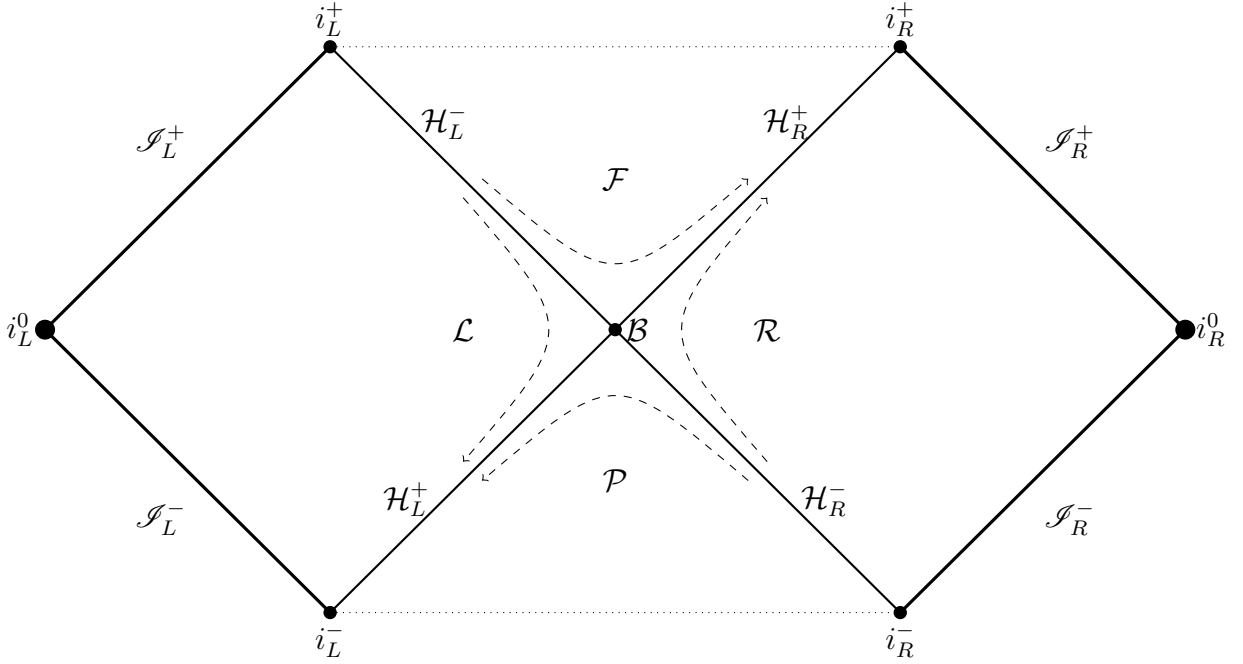

The background metric $g_{ab}$ admits a Killing vector field $\xi^{a}$. The Killing horizon $\mathcal{H}$ is defined as the codimension 1 submanifold of $\mathcal{M}$ on which $g_{ab} \xi^a \xi^b = 0$ with the bifurcation surface $\mathcal{B}$ being the codimension 2 submanifold on which $\xi^a = 0$. Such spacetimes admit a bifurcate Killing horizon structure with two Killing horizons: the \emph{future horizon} ($\mathcal{H}^+$) and the \emph{past horizon} ($\mathcal{H}^-$) intersecting at $\mathcal{B}$. Furthermore $\mathcal{B}$ divides each of them into \emph{left} and \emph{right} horizons: denoted by $\mathcal{H}_L^+$ and $\mathcal{H}_R^+$ for the future horizon, and $\mathcal{H}_L^-$ and $\mathcal{H}_R^-$ for the past horizon. The 3-surface $\hor \cup i_R^+ \cup \mathscr{I}^+$ serves as the surface on which data is defined for bulk field in the spacetime region $\mathcal{P} \cup \mc{R}$ provided it is globally hyperbolic. A generic asymptotically flat spacetime with a bifurcate Killing horizon is depicted in \cref{bif}. We now summarize the assumptions on geometry and linear fields on this spacetime as: 
\begin{enumerate}
    \item The manifold $(\mc P \cup \mc R, g)$ is globally hyperbolic. \label{assump1}
    \item For free (linear) fields compactly supported in $\mathcal{P} \cup \mc{R}$, the data can be independently specified on $\hor$ and $i_R^+ \cup \mathscr{I}_R^+$. Furthermore for the majority of the paper, we shall assume that the data is specified only on $\hor$ and the data is zero on $i_R^+ \cup \mathscr{I}_R^+$. We will relax this condition much later in \cref{rem:news-flux} where we will take data on $\mathscr I_R^+$ too. \label{assump2}
\end{enumerate}

On $\mathcal{H}$, the Killing vector field fails to be affinely parametrized and the failure is measured by the surface gravity $\kappa$ of the horizon:
\begin{equation}
    \xi^b \nabla_b \xi^a = \kappa \xi^a \quad (\text{on } \mathcal{H}) 
\end{equation}
which is constant on $\mathcal{H}$ in general relativity.

The surface of interest for our analysis is the horizon \(\hor = \hor_R \union \hor_L\). Let us denote the affinely parametrized tangent vector field on $\hor$ by $l^a$ with $V$ being the affine parameter. The Killing parameters along \(\xi^a\) are \(v\) on \(\hor_R\) and \(v'\) on \(\hor_L\). These are related to the affine parameter by
\be
    & V = e^{\kappa v} \quad (\text{on } \mathcal{H}_R^+) \quad (-\infty < v <\infty) \\
    & V = -e^{\kappa v'} \quad (\text{on } \mathcal{H}_L^+) \quad (-\infty < v' <\infty)
\ee
and, the affine null tangents are related to the Killing field by
\be
    \xi^{a} = \kappa Vl^{a} \quad (\text{on } \mathcal{H}^+)
\ee\\

Consider a smooth one-parameter family of asymptotically-flat metrics $g_{ab}(\lambda)$ on the spacetime manifold $\mathcal{M}$. The background metric described above will be \(g_{ab} = g_{ab}(\lambda = 0)\). Without loss of generality, we can choose \(\hor\) to be a null surface for all \(\lambda > 0\), see \cite{BHBB}.

Let \(l^a(\lambda)\) be the affine normal to \(\hor\) and let \(q_{ab}(\lambda)\) be the pullback of the spacetime metric \(g_{ab}(\lambda)\) to the null surface \(\hor\). Since \(\hor\) is a null surface \(q_{ab}(\lambda)\) satisfies \(q_{ab}(\lambda) l^b(\lambda) = 0\) and defines a family of degenerate metric on \(\hor\). Thus, it can be equivalently viewed as a Riemannian metric \(q_{AB}(\lambda)\) on the cuts of \(\hor\).  The \emph{second fundamental form} \(K_{ab}\) of $\mathcal{H}^+$ is defined as
\begin{equation}
    K_{ab}(\lambda) = \nabla_{a}(\lambda) l_{b}(\lambda)
\end{equation}
where \(\nabla_a(\lambda)\) is the covariant derivative compatible with \(g_{ab}(\lambda)\). Further the pullback to \(\hor\) of \(K_{ab}(\lambda)\) is orthogonal to \(l^a(\lambda)\), and can be denoted as \(K_{AB}(\lambda)\). Since \(\mc H\) is a null surface, Frobenius' theorem implies that \(K_{AB}(\lambda)\) is symmetric. Decomposition of $K_{AB}(\lambda)$ into a trace and symmetric trace-free parts gives the expansion ($\vartheta$) and shear ($\sigma_{AB}$), respectively, as
\begin{equation} \label{expdef}
    K_{AB}(\lambda) = \frac{1}{2}\vartheta(\lambda) q_{AB}(\lambda) + \sigma_{AB}(\lambda)
\end{equation}
The expansion and shear are related via the vacuum Raychaudhuri's equation for null geodesic congruences:
\begin{equation} \label{Raychaudhuri}
    \partial_V \vartheta(\lambda) = -\frac{1}{2}\vartheta(\lambda)^2 - \sigma_{AB}(\lambda)\sigma^{AB}(\lambda)
\end{equation}
where $V$ is the affine parameter on $\mathcal{H}^+$. Note that for $\lambda = 0$, $\hor$ is a Killing horizon and hence $\vartheta = \vartheta (\lambda = 0) = 0$ and $\sigma_{AB} = \sigma_{AB}(\lambda = 0) = 0$. For more details refer to sec.~9.2 of \cite{Waldbook}. \\

The first and second order metric perturbations over the background are defined by
\be
    \delta g_{ab} \defn \lb. \frac{d g_{ab}(\lambda)}{d\lambda} \rb|_{\lambda=0} \eqsp \delta^2 g_{ab} \defn \lb. \frac{d^2 g_{ab}(\lambda) }{d\lambda^2} \rb|_{\lambda=0}
\ee
We denote the perturbations of other geometric quantities derived from the metric in a similar way by \(\delta\) and \(\delta^2\).

So far \(\hor\) is a null surface for all \(\lambda >0\), but we can also ensure that it is the future horizon of the spacetime to first order in \(\lambda\). To do so note that the diffeomorphism invariance of general relativity implies that the perturbations have the additional gauge freedom
\be\label{eq:gauge-freedom}
    \delta g_{ab} \mapsto \delta g_{ab} + 2 \nabla_{(a} \chi_{b)}
\ee
for any vector field \(\chi^a\) on the manifold \(\mc M\). Under such a linearized diffeomorphism the horizon will also be perturbed. However, it can be shown \cite{BHBB} that using such a linearized diffeomorphism the event horizon of the perturbed spacetime can be taken to coincide with \(\hor\). This implies that we can always choose
\begin{equation}
    \delta \vartheta = 0 \quad (\text{on } \hor)
\end{equation}
We can further also impose (see \cite{HWZ})
\begin{align} 
    \xi^a\delta g_{ab} = 0 \hspace{10pt} (\text{on } \mathcal{H}) \label{gauge1} \\
    \nabla_c(\xi^a\xi^{b}\delta g_{ab}) = 0 \hspace{10pt} (\text{on } \mathcal{H}) \label{gauge2}
\end{align}
Both these conditions can be imposed without loss of generality since they hold for all perturbations in Gaussian null coordinates based on the Killing parameter \(v\), see \cite{HWZ}. Further, \cref{gauge1} ensures that \(\xi^a\) remains the null normal to \(\mc H\) under perturbations and \cref{gauge2} ensures that \(\delta \kappa = 0\) on \(\mc H\).

The condition \cref{gauge1} implies that the pullback of the metric perturbation \(\delta g_{ab}\) to the horizon is orthogonal to \(\xi^a\) and hence \(l^a\). Thus on the horizon, we will denote the data of the metric perturbation \(\delta g_{ab}\) by \(\delta g_{AB}\). The perturbation of the shear is then
\begin{equation} \label{shear1}
    \delta \sigma_{AB} = \frac{1}{2}\partial_V\delta g_{AB}
\end{equation}
The second variation of \cref{Raychaudhuri} relates the second variation of the expansion to the first variation in shear as
\begin{equation} \label{Amal1}
    \partial_V \delta^2\vartheta = -2 \delta\sigma_{AB}\delta\sigma^{AB}
\end{equation}

\section{Classical phase space and quantization of metric perturbations at future horizon}
\label{sec:phase-space}

The Lagrangian \(4\)-form of general relativity equips the theory with a symplectic structure on the space of solutions \cite{Lee-Wald,Iyer-Wald}. The \emph{symplectic structure} is encoded in the symplectic current \(3\)-form \(\omega(g; \delta_1 g, \delta_2 g)\) which is antisymmetric bilinear in the two perturbations and is conserved when \(g_{ab}\) satisfies the equation of motion and \(\delta_{1,2}g_{ab}\) both satisfy the linearized equations of motion. The integral of the symplectic current on any Cauchy surface is thus independent of the Cauchy surface chosen, and defines a symplectic form on the covariant phase space of solutions:
\be\label{eq:symp-bulk}
    W_\Sigma(g; \delta_1g, \delta_2 g) \defn \int_\Sigma d^3x~ \omega(g; \delta_1g, \delta_2 g)
\ee
where \(d^3x\) denotes the induced volume element on the Cauchy surface \(\Sigma\).

To quantize the linearized perturbations, we will also need the Poisson bracket on classical observables, i.e. functions on phase space. However, the symplectic form in general relativity is only \emph{weakly non-degenerate} and the Poisson bracket cannot be defined on arbitrary functions on phase space. A particular class of classical observables of interest for quantization, which do admit a Poisson bracket, are the locally-smeared linearized fields:
\be\label{eq:pert-smeared}
    \delta g(f) \defn \int_{\mc M} d^4x ~ \delta g_{ab} f^{ab}
\ee
where $f^{ab}$ is a symmetric test tensor of compact support which in addition is divergence-free, \(\nabla_a f^{ab} = 0\). Restricting to divergence-free test tensors eliminates the gauge ambiguity under linearized diffeomorphisms \cref{eq:gauge-freedom}.

When smeared with such test tensors, the classical advanced (\(A\)) and retarded (\(R\)) Green's functions are well-defined. Then, the smeared local observables \(\delta g(f)\) can be written as a symplectically-smeared observable \cite{Gautam, Gautam2}
\be\label{eq:local-field}
    \delta g(f) = W_\Sigma(\delta g,  Ef)
\ee
where $E = R-A$ is the \emph{causal propagator}, and \((Ef)_{ab}\) is a solution of the linearized Einstein equations. The Poisson bracket of these observables follows from \cref{eq:local-field} as:
\be \label{eq:poisson-bracket}
    \lb\{ \delta g(f_1), \delta g(f_2) \rb\} = - W_\Sigma(Ef_1, Ef_2)
\ee

For our purposes it will be convenient to rewrite the locally-smeared observable \(\delta g(f)\) in terms of the induced data on the horizon $\hor$ (and there is no data on $i^+_R \cup \scri^+_R$ due to assumption \cref{assump2}). First, the symplectic form \cref{eq:symp-bulk} can be evaluated on the horizon to get \cite{Gautam, Gautam2} 
\be\label{symplectic1}
    W_{\mathcal{H}^+}(g; \delta_{1}g, \delta_2g) = - \frac{1}{8\pi}\int_{\mathcal{H}^+} dV d\Omega_2 (\delta_{1} \sigma^{AB}\delta_2g_{AB} - \delta_1g^{AB}\delta_2 \sigma_{AB})
\ee
where \(\delta g_{AB}\) is the data on \(\hor\) for the perturbation \(\delta g_{ab}\) as explained before in \cref{GeometrySec},

Secondly, to write the locally-smeared field \cref{eq:local-field} on the horizon let \(s_{AB}\) be the data on \(\hor\) for the perturbation \((Ef)_{ab}\). Then we have using \cref{eq:local-field} and \cref{symplectic1}:
\be\label{eq:hor-field}
    \delta\sigma(s) = \int_{\hor}dV d\Omega_2~ \delta\sigma_{AB} s^{AB} = -4\pi \delta g(f) \, .
\ee
Thus, on the horizon the locally-smeared metric perturbation is represented by the smeared perturbed shear tensor. The Poisson bracket among these horizon observables is determined by the symplectic form (obtained by using \cref{symplectic1} in \cref{eq:poisson-bracket} and then identifying the bulk-smeared fields with three-smeared fields using \cref{eq:hor-field}) and obtained to be:
\be \label{Basic}
    \{\delta\sigma(s_1), \delta \sigma(s_2)\} &= 16\pi^2\{\delta g(f_1), \delta g(f_2)\} = -16\pi^2W_{\mathcal{H}^+}(s_1, s_2) \\
    &= \pi \int_{\mathcal{H}^+}dVd\Omega_2 (s_1^{AB}\partial_Vs_{2,AB} - \partial_Vs_{1}^{AB}s_{2,AB})
\ee

The Hamiltonian generator of the Killing flow along \(\xi^a\) on $\hor$ is given by the flux\footnote{Note that \(F_\xi\) in \cref{Flux1} is the flux of the ``modified'' canonical energy of gravitational perturbations defined by Hollands and Wald \cite{BHBB}.}
\be \label{Flux1}
    F_{\xi} \defn W_{\mathcal{H}^+}(\delta g, \Lie_{\xi}\delta g) = \frac{\kappa}{4\pi}\int_{\mc{H}^+}dV d\Omega_2 V \delta \sigma_{AB} \delta \sigma^{AB} 
\ee
This flux generates the Killing flow in the sense that the Poisson bracket between the smeared observables and the flux is given by:
\be \label{Flux}
    \{F_{\xi}, \delta \sigma(s)\} = -W_{\mc{H}^+}(\Lie_{\xi}\delta g, s) = \frac{1}{4\pi}\delta \sigma(\Lie_{\xi} s)
\ee
where the first equality follows from the definition of flux as Hamiltonian generator of the Killing flow and the second equality employed \cref{symplectic1}.

Further for the Killing field $\xi^a$ it can be shown that the symplectic current \(3\)-form satisfies \cite{BHBB}:
\be\label{boundary}
    \omega(g; \delta g, \Lie_{\xi}\delta g) = d[\delta^2Q_{\xi} - \xi\cdot \delta\theta(g, \delta g)]
\ee
where \(Q_\xi\) is the Noether charge \(2\)-form and \(\theta(g,\delta g)\) is the symplectic potential of general relativity (see \cite{Lee-Wald, Iyer-Wald}). Integrating \cref{boundary} over $\hor$ and comparing to \cref{Flux1} gives
\be \label{boundary1}
    F_{\xi} =  \int_{\hor} d[\delta^2Q_{\xi} - \xi\cdot \delta\theta(g, \delta g)]
\ee

Hollands, Wald and Zhang (HWZ) \cite{HWZ} have shown that the pullback of $\theta$ to a Killing horizon is a total variation \(\underline{\theta}_{\mathcal{H}} = \delta B_{\mc{H}}\). Using this, they proposed that the entropy of a dynamical black hole on any cut \(\mc C\) is given by\footnote{The HWZ entropy in \cref{eq:HWZ-entropy} is equivalent to one of the boundary charges on finite null surfaces obtained in \cite{CFP, CP} though it was not identified as an entropy. The extension of the crossed product construction and von Neumann entropy to include the other boundary charges has been investigated in \cite{Gautam2}.} 
\be\label{eq:HWZ-entropy}
    S_{\rm HWZ}[\mc C] \defn \beta \int_{\mc C} (Q_{\xi} - \xi\cdot B_{\mc H}) = \frac{1}{4}A[\mc C] - \frac{1}{4} \int_{\mc C} d\Omega_2~ V \vartheta
\ee
where \(A[\mc C]\) is the area of the cut \(\mc C\) and $\beta = \frac{2\pi}{\kappa}$ with $\kappa$ being the surface gravity of the Killing horizon. Note that the last term containing the expansion \(\vartheta\) vanishes on a Killing horizon and thus \cref{eq:HWZ-entropy} gives the classical black hole entropy. Thus, in our case, in the background spacetime \cref{eq:HWZ-entropy} is just the Bekenstein-Hawking entropy of the black hole, however, the second variation of expansion \(\delta^2\vartheta\) will in general be non-zero and hence the entropy will have the leading order dynamical term at second-order. This entropy is motivated by a ``physical'' process first law where the change in HWZ entropy between two cuts of the horizon equals to the energy fallen through th horizon between the two cuts. As the first law is valid only at leading order (second order in perturbations for linearized gravity), the HWZ entropy is intended to be used only at the leading order \cite{HWZ}.

Using \cref{eq:HWZ-entropy} in \cref{boundary1} we can write the flux \(F_\xi\) as the difference of terms at the boundaries at \(i^+_R\) and \(i^-_L\). To do so we evaluate \(S_{\rm HWZ}[\mc C]\) (from \cref{eq:HWZ-entropy}) on a cut \(\mc C\) with \(V = \text{constant}\) and then take \(V \to \pm \infty\). This gives
\be \label{Fluxop}
    F_{\xi} = \frac{1}{\beta} \delta^2S_{\rm HWZ}[i^+_R] - \frac{1}{\beta} \delta^2S_{\rm HWZ}[i^-_L] 
\ee
where we have assumed that the metric perturbations decay sufficiently fast at asymptotically early and late times such that the limit of the HWZ entropy as $V \to \pm \infty$ is finite. We also note that $\delta^2S_{\text{HWZ}}[\mc C]$ depends on the second order metric perturbation in general and hence is not an observable on the phase space of linearized perturbations. However, the difference of the second order HWZ entropies is the flux of linearized perturbations (as in \cref{Fluxop}) which is an observable.
  
Now equipped with the definitions of classical observables on the covariant phase space as well as the Hamiltonian generating the Killing flow, we will now proceed to quantization of this theory on the horizon. 

\subsection{Quantization on the horizon}
\label{sec:horizon-quant}

We will consider the quantum theory of linearized gravity on this black hole background. On a fixed background spacetime the quantization of the locally-smeared field observables (\cref{eq:pert-smeared,eq:local-field}) can be done using the methods of algebraic QFT in the same manner as for a scalar field; see \cite{FH,HW}. Due to assumption \cref{assump2} we have the bulk-smeared observables equal to appropriately horizon-smeared observables via \cref{eq:hor-field}. Hence we will directly quantize of the data on the horizon given by the shear observables \(\delta\sigma(s)\) in \cref{eq:hor-field}.

Thus, for our quantum field algebra we take the unital \(*\)-algebra $\mathcal{A}_{\mc{H}^+}$ generated by $\op{1}$, $\sigop(s)$ and $\sigop(s)^*$ satisfying the following relations for all test tensors \(s_1^{AB},s_2^{AB}\) on \(\hor\):
\begin{enumerate}
    \item $\sigop(\alpha_1s_1 + \alpha_2s_2) = \alpha_1\sigop(s_1) + \alpha_2\sigop(s_2)$ for all $\alpha_1, \alpha_2 \in \mathbb{R}$;
    \item $\sigop(s)^* = \sigop(s)$;
    \item $[\sigop(s_1), \sigop(s_2)] = -64i\pi^3W_{\mathcal{H}^+}(s_1, s_2)\op{1}$
\end{enumerate}
where the symplectic product is as given in \cref{Basic}.

The \emph{vacuum state} is a Gaussian Hadamard state $\omega_0$ on $\mathcal{A}_{\mc{H}^+}$ with vanishing one-point function and two-point function given by \footnote{This is analogous to the scalar case as mentioned in \cite{Gautam}.}
\be\label{State}
    \omega_0(\sigop_{AB}(x_1)\sigop_{CD}(x_2)) = \frac{1}{\pi}\frac{(q_{A(C} q_{D)B} - \frac{1}{2} q_{AB} q_{CD})\delta_{\mathbb{S}^2}(x_1^A - x_2^A)}{(V_1 - V_2 -i0^+)^2}
\ee
where $x \equiv (V, x^A)$ on $\hor$. It can also be shown that when restricted to acting on right future horizon algebra $\mathcal{A}_{\hor_R}$ the state $\omega_0$ is KMS with respect to Killing flow with inverse temperature $\beta = \frac{2\pi}{\kappa}$ \cite{Gautam, Gautam2}. 

Note that the state $\omega_0$ might not be extendible to a globally (i.e. bulk) KMS state in $\mc R$. While in Schwarzschild spacetime the  Hartle-Hawking state is globally  KMS \cite{HW, Sanders}, it was shown in \cite{Kay} that such states do not exist on Kerr spacetimes. The Unruh state \cite{Unruh1976} for Kerr is not globally KMS but its restriction to the horizon is Hadamard and KMS with 2-point function given by \cref{State}. In our analysis we only require the state \(\omega_0\) to be KMS on the horizon.

Given the algebraic state $\omega_0$ on the algebra $\mathcal{A}_{\hor}$, the GNS construction (see Sec-III.2.2 of \cite{Haag}) gives us a Hilbert space $\mathscr{H}$, a representation of the algebra $\mathcal{A}_{\hor}$ as linear operators on $\mathscr{H}$ and a cyclic separating vector $\ket{ \omega_0 }$ in $\mathscr{H}$ such that if $\op{a} \in \mc{A_{\hor}}$, then $\omega_0(\op{a}) = \braket{ \omega_0 | \op{a} | \omega_0 }$.\\

The flux observable \(F_\xi\) (see \cref{Flux1,Flux}) can be included in the algebra \(\mc A_{\hor}\) as an operator \(\op F_\xi\) satisfying the commutation relations associated to the Poisson bracket \cref{Flux}
\be \label{eq:flux-commutator}
    [\sigop(s), \op F_{\xi}] = -i\sigop(\Lie_{\xi}s)
\ee
The Killing flow along \(\xi^a\) generates a one-parameter family of automorphisms $\alpha_t$ of $\mathcal{A}_{\mc{H}^+}$ by by the unitary action
\be\label{eq:killing-flow}
    \alpha_t[\sigop(s)] = \sigop(s_t) = e^{i\op F_{\xi}t}\sigop(s)e^{-i\op F_{\xi}t}
\ee
where $s^{AB}_t(v, x^A) = s^{AB}(v-t, x^A)$ with $v$ being the Killing time and $x^A$ being coordinates on the cuts of the horizon (which are diffeomorphic to \(\bb S^2\)). The flux operator can be expressed in the form
\be \label{eq:flux-op}
    \op F_{\xi} = \frac{\kappa}{4\pi}\int_{\mc{H}^+}dV d\Omega_2 V :\!\sigop_{AB} \sigop^{AB}\!:~
\ee
where \(:\!\sigop_{AB} \sigop^{AB}\!:\) is the ``normal ordered'' product\footnote{Note that even though the flux operator \(\op F_\xi\) is defined by the integral of a normal ordered product, it is not an element of the Wick algebra (see e.g, \cite{HW}) since it is not smeared with a compact support test function. The locally smeared Wick products on \(\hor\) cannot be defined as operators but only as quadratic forms, i.e., their expectation values in Hadamard states are well-defined but they have infinite fluctuations; see also \cite{IRfinite, Bousso} for similar arguments at null infinity.\label{fn:flux-quad-form}} with respect to the vacuum subtraction given by \cref{State}. 

Finally, we construct a von Neumann algebra of bounded operators on \(\ms H\) which are restricted to the right-wedge component \(\hor_R\) of the horizon (see \cref{bif}). Let $\sigop(s)$ be the smeared perturbed shear operators where the test tensor \(s_{AB}\) is supported on $\hor_R$. The corresponding \emph{Weyl operators} are given by
\be\label{eq:weyl-defn}
    \op{\mc W}(s) \defn \exp(i\sigop(s)) 
\ee
with $\op{\mc W}(s)^* = \op{\mc W}(-s)$. The commutation relations in $\mc{A}_{\hor_R}$ give
\be \label{Weylrelation}
    \op{\mc W}(s_1) \op{\mc W}(s_2) = \op{\mc W}(s_1 + s_2) \exp(8i\pi^2W_{\hor}(s_1, s_2))
\ee
The weak closure of the $C^*$-algebra generated by such Weyl operators is a von Neumann algebra $\mf A(\hor_R, \omega_0)$ (see \cite{vN,Kadison1}). The commutant of this algebra is \footnote{This follows from Haag duality that the commutant of the algebra supported in a subregion is equal to the algebra supported in its casual complement; see \cite{AlanGabriel}}:
\begin{equation}
    \mf A(\hor_R, \omega_0)' = \mf A(\hor_L, \omega_0)
\end{equation}
and hence $\mf A(\hor_R, \omega_0) \cap \mf A(\hor_R, \omega_0)' = \mathbb{C}\op{1}$. Thus $\mf A(\hor_R, \omega_0)$ is a von Neumann factor. Further since \(\mf A(\hor_R, \omega_0)\) is the subalgebra restricted to \(\hor_R\), it can be shown that this is a Type-III factor \cite{Araki}.

Now being a Type-III factor, $\mf A(\hor_R, \omega_0)$ does not have any renormalizable trace and hence any density matrices. Thus, reduced states (represented by density matrices) on the horizon $\hor_R$ and their entropy are both ill-defined. As discussed in the introduction we will make use of the crossed product construction to obtain a new algebra which is a Type-II factor and hence does have a well-defined entropy for states \cite{CLPW,Gautam, Sorce, Witten22}. Moreover we shall see that crossed product comes naturally as we take perturbative gravitational constraint (which will be a quantum version of \cref{Fluxop}) into account similar to the way done in \cite{Gautam}.




\subsection{Tomita-Takesaki modular theory}
\label{sec:TTT}

To construct the Type-II crossed product algebra we recall some machinery from Tomita-Takesaki theory adapted to our setting below.

By the Reeh-Schlieder theorem, the vacuum state $\ket{ \omega_0 }$ is cyclic and separating with respect to the algebra $\mf A(\hor_R, \omega_0)$. Hence, we can apply the tools of Tomita-Takesaki modular theory with the pair $(\mf A(\hor_R, \omega_0), \ket{ \omega_0 })$. The \emph{Tomita operator} $\op S_{\omega_0}$ is defined by
\be
    \op S_{\omega_0}\op a \ket{ \omega_0 } \defn \op a^* \ket{ \omega_0 } \quad \forall \op a \in \mf A(\hor_R, \omega_0)
\ee
The Tomita operator is anti-linear and has polar decomposition:
\be
    \op S_{\omega_0} = \op J_{\omega_0}\op\Delta_{\omega_0}^{\half}  
\ee
where the \emph{modular conjugation} $\op J_{\omega_0}$ is anti-linear with $\op J_{\omega_0}^* = \op J_{\omega_0}$ and the \emph{modular operator} $\op\Delta_{\omega_0}$ is positive definite operator. The corresponding \emph{modular Hamiltonian} is
\be
    \op{H_{\omega_0}} = - \log \op{\Delta_{\omega_0}}
\ee
which generates the \emph{modular flow}
\be\label{eq:modular-flow-defn}
    \op a \mapsto \op\Delta_{\omega_0}^{-it}\op{a}\op\Delta_{\omega_0}^{it} \quad \forall \op{a} \in \mf A(\hor_R, \omega_0)
\ee
By a fundamental theorem of Tomita (proven by Takesaki in \cite{Tak70}), modular flow forms a \(1\)-parameter group of automorphisms of the algebra and, the vacuum state $\ket{\omega_0}$ is KMS with unit inverse temperature with respect to this modular flow.

Now, we have two (automorphic) flows defined on the algebra \(\mf A(\hor_R, \omega_0)\): the Killing flow generated by the flux operator \(\op F_\xi\) through the horizon (\cref{eq:flux-op}) and modular flow generated by the modular Hamiltonian \(\op H_{\omega_0}\) (\cref{eq:modular-flow-defn}). The state $\ket{ \omega_0 }$ is a KMS state with respect to both the Killing flow (with inverse temperature $\beta = \frac{2\pi}{\kappa}$) and the modular flow (with inverse temperature \(1\)). It follows from the \emph{Bisognano-Wichmann theorem} \cite{BW, Sewell, Verch} that the two flows on the algebra can be identified such that the generators are related by 
\be \label{BWT}
    \op H_{\omega_0} = \beta\op F_{\xi}
\ee

Let $\ket{ \omega } \in \mathscr{H}$ be another cyclic separating state. One can now define the \emph{relative Tomita operator} $\op S_{\omega|\omega_0}$ as:
\be
    \op S_{\omega|\omega_0}\op{a} \ket{ \omega_0 } = \op{a}^* \ket{ \omega } \quad \forall \op{a} \in \mf A(\hor_R, \omega_0) 
\ee
The relative Tomita operator has an analogous polar decomposition in terms of the \emph{relative modular conjugation} \(\op J_{\omega|\omega_0}\), the \emph{relative modular operator} \(\op \Delta_{\omega|\omega_0}\) and the \emph{relative modular Hamiltonian} as
\be
    \op S_{\omega|\omega_0} = \op J_{\omega|\omega_0} \op \Delta_{\omega|\omega_0}^{\half} \eqsp \op{H_{\omega|\omega_0}} = - \log \op\Delta_{\omega|\omega_0}
\ee
Finally, Araki's \emph{relative entropy} is defined by:
\be\label{eq:rel-entropy-defn}
    S(\omega|\omega_0) \defn - \braket{ \omega | \op H_{\omega|\omega_0} | \omega }
\ee
We emphasize here that the relative entropy is defined for any von Neumann factor for any two (cyclic and separating) states. In contrast the von Neumann entropy of a state is ill-defined for a Type-III factor due to lack of a renormalized trace and renormalizable density matrices.

\subsection{Dressed observables and crossed product algebra} \label{dressing}

Using the machinery of modular theory, we now recall construction of the crossed product algebra and the corresponding trace (see \cite{CLPW, Gautam}). This will lead us to the entropy of the reduced density matrix corresponding to QFT coupled with an observer (that we will explain shortly) on the right future horizon.

In full quantum gravity observables are expected to be diffeomorphism invariant. However, since we are only considering perturbative gravity off of a background with Killing symmetry, we should consider observables which are invariant under the Killing flow. However, local observables are not invariant under any diffeomorphism, in general as can be seen readily in \cref{eq:killing-flow}. It has been suggested that suitably defined ``gravitational dressed'' or ``relational'' observables can be constructed which are invariant \cite{DeWitt1962Quantization, DeWitt1967Canonical, DonnellyGiddings2016, DonnellyFreidel2016, BrownKuchar1995Dust, CiambelliFreidelLeigh2023, GoellerHoehnKirklin2022, Rovelli1996Relational, Dittrich2006, Tambornino2012}.

In \cite{CLPW} such observables were defined by introducing an additional ``observer degree of freedom'' which couples to the QFT in the spacetime. The joint ``relational observables'' are then invariant under the Killing symmetry of the background spacetime. This formalism has been extended to asymptotically-flat black hole spacetimes where instead of an observer one ``dresses'' to the the gravitational charges at the boundaries of the spacetime under consideration \cite{Gautam, Gautam2}. The addition of such observers or charges into the operator algebra naturally leads to the crossed product construction, which yields an extended algebra which is a Type-II factor and has a good definition of entropy of states. We now define such observables for perturbative quantum gravity.

Recall that the algebra elements of $\mf A(\hor_R, \omega_0)$ are not invariant under the Killing flow. In particular, from \cref{eq:killing-flow,eq:weyl-defn}, the Weyl operators \(\op{\mc W}(s)\) transform as 
\be\label{eq:weyl-killing-flow}
    \alpha_t[\op{\mc W}(s)] = \op{\mc W }(s_t) = e^{i\op F_{\xi}t} \op{\mc W}(s) e^{-i\op F_{\xi}t}
\ee
To construct observables invariant under the Killing flow, we will need to suitably ``dress'' the Weyl operators \(\op{\mc W}(s)\) with the gravitational charges. For linearized gravity, such charges are given by the limiting values of the second variation of the HWZ-entropy (see \cref{Fluxop}). However, since these charges involve a second variation they are not included in our algebra of observables a priori. Thus, we extend our operator algebra to include an additional one-dimensional degree of freedom as follows. Consider
\be
    X \defn \frac{1}{\beta} \delta^2 S_{\rm HWZ}[i^+_R]
\ee
This quantity \(X\) will serve as the analogue of the ``observer Hamiltonian'' as we explain below. We represent \(X\) as a multiplication operator \(\op X\) on the Hilbert space \(L^2(\bb R)\) of square-integrable functions \(f(X)\). Thus our extended Hilbert space will be
\be
    \Hilb^{\rm ext} \defn \Hilb \otimes L^2(\bb R)
\ee
We now use \cref{Fluxop}, to then represent \(C \defn \frac{1}{\beta} \delta^2 S_{\rm HWZ}[i^-_L]\) as an operator on \(\Hilb^{\rm ext}\) by
\be \label{globalconstraint}
    \op C = \op X - \op F_{\xi}
\ee
Since \(\op C\) lies in the causal complement of the right wedge \(\mc R\) it commutes with all ``physical'' observables in \(\mc R\). Thus \cref{globalconstraint} acts as a ``gravitational constraint'' on the physical observables in \(\mc R\). The observables which commute with \(\op C\) will be our ``dressed observables''.

To construct these observable explicitly, we introduce the ``canonical conjugate'' operator \(\op p\) on \(L^2(\bb R)\) satisfying\footnote{\label{fn:povm}In our case, the spectrum of the operator \(\op X\) will not be the entire real line and thus \(\op p\) cannot be a self-adjoint operator on the observer Hilbert space. This can be handled rigorously using \emph{positive operator-valued measures} (POVM), see e.g. \cite{FJLRW}. This subtlety will not affect our main results.}
\be\label{time}
    [\op X, \op p] = i \1
\ee
Then we extend the Killing flow with $-\op X$ generating the Killing flow in the $L^2(\mathbb{R})$, i.e. if $\op{b}$ is a linear operator in $L^2(\mathbb{R})$, then
\be\label{eq:ext-killing-flow}
    \alpha_t[\op{b}] = e^{-i\op{X}t}\op{b}e^{i\op{X}t}
\ee
so that $\alpha_t[\op{X}] = \op{X}$ and $\alpha_t[\op{p}] = \op{p} + t\op{1}$. Thus, \(\op p\) acts as the ``time operator'' conjugate to the ``observer Hamiltonian'' \(- \op X\). Thus $-\op C$ acts as the total Hamiltonian of the joint QFT and ``observer'' system.

The operators that commute with the constraint \(\op C\) are then obtained as follows. Consider the Weyl operators (\cref{eq:weyl-defn}) ``dressed'' by the flux operator and the ``observer time'' as\footnote{Note that \(e^{-i \op F_\xi \op p}\) is a well-defined operator on \(\Hilb^{\rm ext}\) since we can ``diagonalize'' \(\op p\) on \(L^2(\bb R)\);  see also \cref{fn:povm}.}
\be\label{eq:weyl-dress-defn}
    \op{\mc W}(s, \op p) \defn e^{-i \op F_\xi \op p} \op{\mc W }(s) e^{i \op F_\xi \op p}
\ee
From the definition of dressed Weyl operators (\cref{eq:weyl-dress-defn}) and using \cref{time} and \cref{eq:flux-commutator}, we have:
\be
    [\op{X}, \op{\mc W}(s, \op{p}) ] = i \op{\mc W}(\Lie_\xi s, \op p) \eqsp [\op p, \op{\mc W}(s, \op p)] = 0
\ee
Further, \cref{eq:weyl-killing-flow,eq:ext-killing-flow} imply that under the Killing flow the dressed Weyl operators are invariant, i.e., 
\be
    \alpha_t[\op{\mc W}(s, \op p)] = \op{\mc W}(s, \op p)
\ee
and thus commute with the constraint (\cref{globalconstraint}). We can similarly ``dress'' all the operators in \(\mf A(\hor_R, \omega_0)\) to get a subalgebra of all bounded operators on \(\mathscr{H}\otimes L^2(\mathbb{R})\) which are invariant under the Killing flow. This defines the \emph{dressed algebra} $\mf A^{\text{dress}}(\hor_R, \omega_0)$ as
\be
    \mf A^{\text{dress}}(\hor_R, \omega_0) = \lb\{ e^{-i\op F_{\xi} \op p} \op{a} e^{i\op F_{\xi} \op p } ~|~ \op{a} \in \mf A(\hor_R, \omega_0) \rb\}
\ee
Since these dressed observables are obtained by just conjugation with a unitary operator, this dressed algebra is isomorphic to the original field algebra
\be
    \mf A^{\text{dress}}(\hor_R, \omega_0) \cong \mf A(\hor_R, \omega_0)
\ee
and hence $\mf A^{\text{dress}}(\hor_R, \omega_0)$ is still a Type-III factor. Now define the \emph{extended algebra} $\mf A^{\text{ext}}(\hor_R, \omega_0)$ as the \emph{double commutant} of the algebra generated by the dressed Weyl operators and the operator \(\op X\) on the ``observer'' Hilbert space \(L^2(\bb R)\):
\be
    \mf A^{\text{ext}}(\hor_R, \omega_0) = \{e^{-i \op F_\xi \op{p}} \op{a} e^{i \op F_\xi \op{p}}, \op{X} \hspace{5pt} | \op{a} \in \mf A(\hor_R, \omega_0)\} ''
\ee
Using \cref{BWT} we see that the dressing with the flux operator is equivalent to dressing with the modular Hamiltonian \(\op H_{\omega_0}/\beta\) and this construction is also called the \emph{crossed product algebra} of $\mf A^{\text{dress}}(\hor_R, \omega_0)$ with its modular automorphism group and is precisely the subalgebra of bounded operators in $\Hilb^{\rm ext}$ commuting with the charge $\op C$ \cite{CLPW, FJLRW, Gautam, Sorce}. By Takesaki's theorem \cite{Tak70}, $\mf A^{\text{ext}}(\hor_R, \omega_0)$ is a Type-II von Neumann factor.

Since the extended algebra is a Type-II factor, one can define \emph{renormalized traces}\footnote{A renormalized trace is a \emph{tracial weight} which is finite for all finite projections in the algebra; see \cite{Sorcevon,Kadison2}.} and density matrices on the extended algebra. For Type-$\text{II}_1$ factors, the identity operator is trace-class and the trace is normalized by $\Tr (\op{1}) = 1$. However as we shall show shortly, in our case the extended algebra will be Type-$\text{II}_{\infty}$, where the identity is not trace-class resulting in the trace to be ambiguous up to scale.

On $\mf A^{\text{ext}}(\hor_R, \omega_0)$ we use the renormalized trace given by (see \cite{CLPW, Gautam})
\be \label{Trace}
    \Tr (\op{a}) = \beta \int_{\mathbb{R}}dX e^{\beta X} \braket{ \omega_0, X | \op{a} | \omega_0, X}
\ee
where $\op{a} \in \mf A^{\text{ext}}(\hor_R, \omega_0)$, $\ket{\omega_0, X} \defn \ket{ \omega_0 } \otimes \ket{ X }$ and \(\op X \ket{X} = X \ket{X}\). For the case of an observer moving in a worldline in de Sitter spacetime, considered in \cite{CLPW, FJLRW}, the extended algebra is a Type-$\text{II}_1$ factor and if the spectrum of $\op{X}$ is $(-\infty, 0)$ then the trace \cref{Trace} is normalized. In our case, we will obtain the spectrum of $\op X$ in the following way: 

We assume that the metric perturbation decays fast enough that the symplectic flux through $i^+_R$ vanishes.\footnote{The vanishing of the symplectic flux through \(i^+_R\) has been shown for massless fields on Schwarzschild spacetime in \cite{Moretti}. For gravitational perturbations such decay results should follow from linearized stability of black holes, but we have not investigated this in detail.} Hence, the boundary charge \(X\) at $i_R^+$ obtained by taking the limit along $\hor_R$ equals the one obtained from the limit along $\mathscr I_R^+$. From, \cref{assump2}, no radiation flux on $\mathscr I^+_R$ implies that \(X\) equals the second variation of the ADM Hamiltonian associated with the symmetry \(\xi^a\) at spatial infinity\footnote{In general \(\xi^a\) is a linear combination of a time translation and spatial rotation at spatial infinity. Thus, the corresponding ADM Hamiltonian contains both the ADM mass/energy and the ADM angular momentum.} 
\begin{equation} \label{crucial}
    \op X = \op{\delta^2 H}_R^{\rm ADM}
\end{equation}
In general, the second variation of the ADM Hamiltonian need not be bounded from above. Thus in our case, we have $\text{Tr}(\op 1) = \infty$ and hence the extended algebra is $\mf A^{\text{ext}}(\hor_R, \omega_0)$ is Type-$\text{II}_{\infty}$. This is a well known fact for Schwarzschild black hole in asymptotically flat spacetimes \cite{CLPW}. We will use the trace \cref{Trace} and the scaling ambiguity in the trace will show up as a state-independent additive ambiguity in the von Neumann entropy (\cref{Master} below).\\

\begin{remark}[No maximal entropy state]
\label{rem:max-entropy}
Since the algebra $\mf A^{\text{ext}}(\hor_R, \omega_0)$ is Type-$\text{II}_{\infty}$ there does not exist any state which maximizes the von Neumann entropy. In fact, it can be shown that for Type-$\text{II}_{\infty}$ factors, there exist density matrices with arbitrarily large trace --- the trace for finite projections in Type-$\text{II}_{\infty}$ factors takes value in the entire range $[0, \infty)$, see sec~8.5 of \cite{Kadison2}.
\end{remark}

\section{Entropy of classical-quantum coherent states of linearized quantum gravity} \label{EntropySec}

With the tools of modular theory and the crossed product construction at hand, we will now obtain the entropy of states in perturbative gravity on the horizon. 

Consider a classical metric perturbation \(h_{ab}\) in the right wedge $\mathcal{R}$ and let $h_{AB}$ be the corresponding data on \(\hor_R\). In the quantum field theory, this metric perturbation can be represented by a coherent state defined as follows. For any perturbation \(h_{AB}\) on \(\hor_R\), define the unitary
\be
    \op{U} \defn \exp(-i\sigop(h)/16\pi^2) 
\ee
The (algebraic) coherent state on the right future horizon corresponding to the metric perturbation \(h_{AB}\) is then given by
\be\label{coherent}
    \omega_h(\op{a}) \defn \omega_0(\op{U^*aU}) \quad \forall \op{a} \in \mathfrak A(\hor_R, \omega_0)
\ee
From the above definition it follows that the 1-point function of this coherent state is $\omega_h(\sigop(s)) = \int dV d\Omega_2~ \delta\sigma_{AB} s^{AB}$ where $\delta\sigma_{AB} = \frac{1}{2}\partial_V h_{AB}$ is the shear of the perturbation and hence, this coherent state is indeed the perturbed state of the black hole. \Cref{coherent} only defines the algebraic state $\omega_h$ on observables supported on $\hor_R$. Hence it does not have a unique representation in the GNS Hilbert space $\mathscr H$ constructed from the algebra supported on the entire horizon $\mathcal{A}_{\hor}$. However, \(\omega_h\) has an unique representation \(\ket{\omega_h}\) in the \emph{natural cone} $\mathcal{P}^{\#}$ of $(\mathfrak A(\hor_R, \omega_0), \ket{ \omega_0 })$ as we describe next.

The modular conjugation \(\op J_{\omega_0}\) (see \cref{sec:TTT}) defines a map from the algebra $\mf A(\hor_R, \omega_0)$ to its commutant $\mf A(\hor_R, \omega_0)' = \mf A(\hor_L, \omega_0)$:
\be \label{eq:j-map}
    j_{\omega_0}: \mf A(\hor_R, \omega_0) \to \mf A(\hor_R, \omega_0)' : 
    \op a \mapsto j_{\omega_0}(\op{a}) = \op J_{\omega_0}\op{a}\op J_{\omega_0} 
\ee
The natural cone in $\mathscr{H}$ of $(\mathfrak A(\hor_R, \omega_0), |\omega_0\rangle)$ is defined as the closure:
\be \label{naturalcone}
    \mathcal{P}^{\#} \defn \overline{\{\op{a} j_{\omega_0}(\op{a}) | \ket{ \omega_0} ~|~ \op{a} \in \mf A(\hor_R, \omega_0)\}} 
\ee
where the closure is taken with respect to norm topology on the Hilbert space $\mathscr{H}$. We refer to sec.~2.5.4 of \cite{BR1} for properties of state vectors in the natural cone. An important property is that any normal state, i.e., continuous state with respect to the weak \(*\)-topology on $\mf A(\hor_R, \omega_0)$, has a unique representation in natural cone.

The algebraic state \(\omega_h\) \cref{coherent} is a normal state and thus has an unique representation \(\ket{\omega_h}\) in the natural cone $\mathcal{P}^{\#}$ given by
\be \label{uniquerep}
    \ket{ \omega_h} = \op{U} j_{\omega_0}(\op{U}) \ket{\omega_0}
\ee
where \(j_{\omega_0}\) is the map defined in \cref{eq:j-map}. 
It can be verified that $\ket{ \omega_h}$ is indeed a representation of $\omega_h$ on the algebra $\mf A(\hor_R, \omega_0)$, since we can compute the expected value of any $\op a \in \mf A(\hor_R, \omega_0)$ as
\be
    \braket{ \omega_h | \op a | \omega_h } &= \braket{ \omega_0 | j_{\omega_0}(\op U^*)\op U^* \op a \op U j_{\omega_0}(\op U) | \omega_0 } \\
    &= \braket{ \omega_0 | j_{\omega_0}(\op U^*)j_{\omega_0}(\op U)\op U^* \op a \op U \op |\omega_0 } \\
    &= \braket{ \omega_0|\op U^* \op a \op U \op |\omega_0 } \\
    &= \omega_h(\op a)
\ee
where in the first line we used the fact that $\op J_{\omega_0}^* = \op J_{\omega_0}$ and hence $j_{\omega_0}(\op U)^* = j_{\omega_0}(\op U^*)$. While going to the second line, we used the fact that $j_{\omega_0}(\op U)$ is in the commutant algebra and hence we commuted $j_{\omega_0}(\op U) \in \mf A(\hor_L, \omega_0)$ across operators $\op U, \op U^*, \op a \in \mf A(\hor_R, \omega_0)$. Then in the third equality we used $\op J_{\omega_0}^2 = \op 1$ and hence $j_{\omega_0}(\op U^*) j_{\omega_0}(\op U) = \op 1$.

Corresponding to the coherent state \(\ket{\omega_h}\), we define a ``classical-quantum state'' \(\ket{\hat \omega_h}\) in the extended Hilbert space $\mathscr{H} \otimes L^2(\mathbb{R})$ by
\be \label{state}
    \ket{\hat\omega_h} \defn \int_{\mathbb{R}}dX f(X) \ket{ \omega_h} \otimes \ket{X}
\ee
where $f \in L^2(\mathbb{R})$ which can be interpreted as the ``wavefunction of the observer/boundary charge''.

Since the extended algebra $\mf A^{\text{ext}}(\hor_R, \omega_0)$ is a Type-II factor, there exists a (renormalized) density matrix $\op \rho_{\hat{\omega}_h} \in \mf A^{\text{ext}}(\hor_R, \omega_0)$ corresponding to the state $\ket{\hat{\omega}_h}$ such that
\be \label{density}
    \Tr (\op \rho_{\hat{\omega}_h}\op a) = \langle\hat{\omega}_h|\op a|\hat{\omega}_h\rangle \quad \forall \op a \in \mf A^{\text{ext}}(\hor_R, \omega_0) \eqsp \Tr (\op\rho_{\hat{\omega}_h}) = 1
\ee
where the trace functional $\text{Tr}$ is defined on $\mf A^{\text{ext}}(\hor_R, \omega_0)$ in \cref{Trace}.
This density matrix $\op\rho_{\hat{\omega}_h}$ serves as an analogue of the reduced density matrix of the full semi-classical coherent state on the entire future horizon to the right future horizon. Note that this reduced density matrix is not just a reduced QFT state to $\hor_R$ (which is undefined) but a joint reduced state of the QFT and observer on the right horizon. \\\\

Now, using the trace functional \cref{Trace} further, we can compute the von Neumann entropy of the density matrix $\op\rho_{\hat{\omega}_h}$ as
\be \label{eq:vN}
    S_{\text{vN}}(\op\rho_{\hat{\omega}_h}) = -\text{Tr}(\op\rho_{\hat{\omega}_h} \log \op\rho_{\hat{\omega}_h})
\ee
The computation is tedious which we detail in \cref{appendix:entropy-calc}. Here we will directly use the final result for $S_{\text{vN}}(\op\rho_{\hat{\omega}_h})$. In the approximation that the ``observer wavefunction'' \(f\) is ``slowly varying'' (roughly speaking, the observer wave function's Fourier transform is sharply peaked at zero momentum, see \cref{appendix:entropy-calc}), we have:
\be\label{Master}
    S_{\text{vN}}(\op\rho_{\hat{\omega}_h}) = -S(\omega_h|\omega_0) + \beta \braket{ \op X }_{\hat{\omega}_h} + S(f) + \log\beta
\ee
where $S(f) = -\int_{\mathbb{R}}dX|f(X)|^2\log|f(X)|^2$. This result matches with \cite{Gautam} where a similar relation was derived in the context of scalar free field in a black hole spacetime. \\

The relative entropy between a coherent state and a vacuum state has been is computed in \cite{HI, GautamInfo}. First we note that \cite{HI} computed $S(\omega_0|\omega_h)$ instead of $S(\omega_h|\omega_0)$. However for $\ket{ \omega_h }$ in the natural cone, it can be shown that they are equal. Secondly, \cite{HI} uses the modular operator in the von Neumann algebra generated by Weyl operators supported in the future of some cut with \(V = V_0\), while \cite{GautamInfo} uses the past of some cut. Thus, their expression for relative entropy includes the affine time \(V_0\) of this cut. The corresponding modular flow then corresponds to a ``dilation'' which keeps $V_0$ fixed. In our case, the relevant algebra is the one supported on \(\hor_R\) and \(V_0 = 0\). Thus the relative entropy in our case is
\be\label{relentropy}
    S(\omega_h|\omega_0) = \beta F_{\xi}[\hor_R] = \frac{1}{2}\int_{\hor_R}dV d\Omega_2 V(\delta\sigma_h)^2
\ee
where $\delta\sigma_h = \frac{1}{2}\partial_Vh$ is the perturbed shear at $\hor$ corresponding to $h_{AB}$ and hence $F_{\xi}[\hor_R]$ is the classical radiation flux of the linearized perturbation falling into the black hole through $\hor_R$.

Further, since we have assumed that there is no flux of (perturbed) radiation through null infinity, we have \(\braket{\op X}_{\hat \omega_h} = \braket{\op{ \delta^2 H}^{\rm ADM}_R}_{\hat \omega_h}\), i.e., the second order variation of the ADM Hamiltonian at spatial infinity (associated to the Killing field \(\xi^a\)) also viewed as a quadratic form. Then, \cref{Master} can be written as
\be\label{eq:first-law-type}
    S_{\text{vN}}(\op\rho_{\hat{\omega}_h}) = \braket{ \op{\delta^2 H}^{\rm ADM}_R}_{\hat \omega_h} - \beta F_{\xi}[\hor_R] + S(f) + \log\beta
\ee
Dividing by \(\beta\), the left-hand side is the change in entropy and the right-hand side represents the change in the ``energy'' relative to \(\xi^a\) due to the gravitational perturbation (up to the state-independent constants). Thus, \cref{eq:first-law-type} can be interpreted as a thermodynamic first law for quantized gravitational perturbations.

Next, we relate \cref{Master} to the Hollands-Wald-Zhang entropy for dynamical black holes \cite{HWZ}.
Consider some cut \(\mc C\) of the horizon. Then we can integrate \cref{boundary1} from \(\mc C\) to \(i^+_R\) and use \cref{eq:HWZ-entropy} to get:  
\be \label{Master2}
    \op{X} &= 
    \frac{1}{\beta}\op{\delta^2 S}_{\rm HWZ}[\mc C] + \op F_\xi[{\hor_{\geq \mc C}}]
\ee
where $\op F_\xi[{\hor_{\geq \mc C}}]$ measures the flux through $\mc{H}_{\geq \mc{C}}$, the part of the horizon $\hor_R$ between the section $\mc C$ and $i_R^+$. In general, the truncated flux $\op F_\xi[{\hor_{\geq \mc C}}]$ is a quadratic form and and not operator (see \cref{fn:flux-quad-form}) on $\mathscr H$, and hence, the (quantum) entropy $\op{\delta^2 S}_{\rm HWZ}[\mc C]$ is also a quadratic form on $\mathscr H^{\rm ext}$. Now, using \cref{Master2} in \cref{Master} we get:
\be \label{Master3}
    S(\op\rho_{\hat{\omega}_h}) &= -S(\omega_h|\omega_0) +  \braket{ \op{\delta^2 S}_{\rm HWZ}[\mathcal{C}]}_{\hat{\omega}_h} + \beta \braket{\op F_\xi[\hor_{\geq \mc C}]}_{\hat{\omega}_h} + S(f) + \log\beta
\ee
It can be shown that for a coherent state \(\omega_h\) the expected value \(\braket{\op F_\xi[{\hor_{\geq \mc C}}]}_{\hat{\omega}_h}\) evaluates to the classical flux of the perturbation \(h_{AB}\) through the region \(\hor_{\geq \mc C}\). Using this with, \cref{relentropy,Master3} we have:
\be \label{eq:HWZ-comparison}
    S(\op\rho_{\hat{\omega}_h}) = \braket{ \op{\delta^2 S}_{\rm HWZ}[\mathcal{C}]}_{\hat{\omega}_h} - \beta F_{\xi}[\hor_{<\mc C}] + S(f) + \log\beta
\ee
where 
\be
    F_{\xi}[\hor_{<\mc C}]  = \frac{\kappa}{4\pi}\int_{\hor_{< \mc C}}dV d\Omega_2 V(\delta\sigma_h)^2 
\ee
is the classical flux through $\hor_{< \mc C}$, the part of the horizon $\hor_R$ between the bifurcation surface $\mc B$ and the cut $\mc C$. Thus, at leading order in perturbation theory (for linearized gravity, it is second order), the von Neumann entropy can be related the HWZ entropy of a dynamical black hole at some arbitrary cut $\mc C$ of $\hor_R$ and a flux term measuring the gravitational flux fallen into the black hole till the cut $\mc C$.\\

\begin{remark}[Gravitational flux through null infinity]
\label{rem:news-flux}

In the main analysis of the paper, for simplicity we assumed that the metric perturbation does not radiate through null infinity. However, we can relax this assumption and account for radiation at \(\scri^+_R\). In asymptotically-flat spacetimes the linearized radiative degree of freedom at null infinity is the perturbed News tensor which can be quantized following \cite{Ashtekar1981, Ashtekar1987, IRfinite}. The algebra generated by these News operators admits a vacuum state which is invariant under the Bondi-Metzner-Sachs (BMS) group.\footnote{Note that neither the Hartle-Hawking nor the Unruh state restricted to \emph{future} null infinity give the BMS-invariant vacuum state.} The corresponding GNS representation gives a Hilbert space representation of the asymptotic radiative operator algebra and the von Neumann algebra of bounded operators on this Hilbert space can be shown to be a factor of type $\text{I}_{\infty}$. For the details in the linearized gravity case we refer to \cite{Gautam2}, we only summarize the changes to our analysis below.

 The radiative data of any asymptotically flat metric perturbation can be represented by a coherent state \(\ket{\psi}\) similar to \cref{coherent}. As in our main analysis we use \(\op X\) as the ``observer'' Hamiltonian and construct the crossed product and the extended Hilbert space on the horizon. Then we take the tensor product \(\ket{\Psi} \defn \ket{\hat\omega_h} \otimes \ket{\psi}\) to represent the full state on the horizon (including the ``observer wavefunction'') and null infinity. Since we have tensored the state \(\ket{\psi}\) at null infinity with the crossed product state \(\ket{\hat\omega_h}\) on the horizon,  \cref{Master} remains unchanged and only the horizon state contributes to the right-hand side.

The horizon Killing field $\xi^a$ of the background spacetime can be shown to be an element of some Poincar\'e subalgebra of the asymptotic BMS symmetry algebra; in general \(\xi^a\) will be a linear combination of a time translation and a rotation in the BMS Lie algebra \cite{AX}. The flow along this vector field on the asymptotic fields is the generated by a flux operator, which we denote by \(\op F_\xi[\scri^+_R]\); the explicit expressions for the flux can be found by linearizing the formulae in, e.g., \cite{GPS,memory-orbits}. This flux through null infinity enters when relating \(\op X\) to the ADM Hamiltonian via
\be
    \op X = \op{\delta^2 H}_R^{\rm ADM} + \op F_{\xi}[\scri_R^+]
\ee
Since the perturbation is represented by a coherent state on null infinity, the expected value of the flux is given by the classical flux expression for the perturbation and \cref{eq:first-law-type} is modified to
\be \label{radiationtakenintoaccount}
    S_{\rm vN}(\op \rho_{\hat{\omega}_h}) = \beta \braket{ \op{\delta^2 H}_R^{\rm ADM}}_{\!\Psi} - \beta F_{\xi}[\hor_R] + \beta F_{\xi}[\scri_R^+] + S(f) + \log \beta
\ee
\end{remark}

\begin{remark}[Matter fields]
\label{rem:matter}
The entire analysis in the paper is unchanged if semi-classical matter is taken into account. The states will now incorporate matter perturbations as tensor products with linearized gravity states. For massless matter fields the flux terms at both the horizon and null infinity will get an additional contribution due the matter fields. For massive fields, there will be no contribution at null infinity but the ``boundary charge'' \(X\) at \(i^+_R\) will be modified by contributions from the massive massive fields (see, for instance, \cite{IRfinite}).
\end{remark}

\begin{remark}[Memory and soft radiation]
\label{rem:memory}
In our main analysis we have ignored the effects of any ``soft radiation'' or the memory effect on both the horizon and at null infinity. However, it now well established that such low frequency radiation plays a crucial role in the quantum theory both at null infinity \cite{Ashtekar1987,IRfinite,amplitudes-soft-thm,memory-orbits} and the horizon \cite{Danielson2022, Danielson2023, Danielson2024, Gralla2023, GautamInfo}. The formalism of Tomita-Takesaki theory has recently been generalized to include such ``soft radiation'' in \cite{GautamInfo}. Using this formalism, we expect that our analysis can be extended to account for such linearized perturbations with non-trivial memory on the horizon and null infinity.
\end{remark}





\acknowledgements
AM is supported by the Visiting Student Program (VSP) at the Raman Research Institute.

\appendix

\section{Calculation of von Neumann entropy for classical-quantum coherent states on the horizon}
\label{appendix:entropy-calc}

In this appendix we details the computations leading to the the expression for the von Neumann entropy (\cref{Master}) using the renormalized trace (\cref{Trace}). 

Using the fact that \(\ket{\omega_h} \in \mc P^{\#}\), we use the trace expression \cref{Trace} in \cref{density}, to get \cite{Gautam, Sorce}:
\be \label{statereduced}
    \op\rho_{\hat{\omega}_h} = \frac{1}{\beta} e^{-i\frac{\op H_{\omega_0}}{\beta}\op{p}} f\lb( \op{X} + \frac{\op H_{\omega_0}}{\beta} \rb) e^{-\frac{\beta\op{X}}{2}} \op\Delta_{\omega_h|\omega_0} e^{-\frac{\beta\op{X}}{2}} f^*\lb(\op{X} + \frac{\op H_{\omega_0}}{\beta} \rb) e^{i\frac{\op H_{\omega_0}}{\beta}\op{p}}
\ee
The derivation of \cref{statereduced} is computationally involved and it involves computing the modular operator for the state $|\hat{\omega}_h\rangle$ in the crossed product factor $\mf A^{\rm ext}(\hor_R, \omega_0)$, denoted by $\op \Delta_{\hat{\omega}_h}$. Then one can seen that $\op \Delta_{\hat{\omega}_h}$ fatorizes into an element of the algebra $\mf A^{\rm ext}(\hor_R, \omega_0)$ and an element of its commutant algebra $\mf A^{\rm ext}(\hor_R, \omega_0)'$ in $\Hilb^{\rm ext}$, i.e. $\op \Delta_{\hat{\omega}_h} = \op\rho_{\hat{\omega}_h} (\op\rho_{\hat{\omega}_h}')^{-1}$ where $\op\rho_{\hat{\omega}_h} \in \mf A^{\rm ext}(\hor_R, \omega_0)$ is identified with the density matrix corresponding to the state $|\hat{\omega}_h\rangle$ in $\mf A^{\rm ext}(\hor_R, \omega_0)$. We shall not show the details of the construction which can be found in Appendix~E of \cite{Sorce}. 

Now note that $f(\op{X} + \frac{\op H_{\omega_0}}{\beta})$ and $ \op\Delta_{\omega_h|\omega_0}$ do not commute:
\be \label{commutator}
    \Big[f(\op{X} + \frac{\op H_{\omega_0}}{\beta}), \op\Delta_{\omega_h|\omega_0}\Big] = f'(\mathbf X)\Big[\frac{\op H_{\omega_0}}{\beta}, \op\Delta_{\omega_h|\omega_0}\Big] + \frac{1}{2}f''(\mathbf X) \Big[\frac{\op H_{\omega_0}^2}{\beta^2}, \op\Delta_{\omega_h|\omega_0}\Big] + \ldots
\ee
and hence the logarithm of $\op\rho_{\hat{\omega}_h} $ in \cref{statereduced} will not distribute additively over the terms in the R.H.S. However, here we make a simplifying assumption that the function $f$ is slowly varying, i.e. derivatives of $f$ are vanishingly small (more precisely the derivatives are much smaller than the corresponding commutators in the \cref{commutator}). In this slowly varying approximation, we have:
\be
    \lb[f \lb( \op{X} + \frac{\op H_{\omega_0}}{\beta} \rb), \op\Delta_{\omega_h|\omega_0} \rb] \approx 0
\ee
Thus now we can take a logarithm of \cref{statereduced} easily as the log will distribute additively over all the commuting operators and we have:
\be\label{comp2}
    -\log{\op\rho_{\hat{\omega}_h}} = e^{-i\frac{\op H_{\omega_0}}{\beta}\op{p}} \lb[ \log\beta  + \op H_{\omega_h|\omega_0} + \beta\op{X} - \log \abs{ f \lb(\op{X + \frac{\op H_{\omega_0}}{\beta}} \rb) }^2 \rb] e^{i\frac{\op H_{\omega_0}}{\beta}\op{p}}
\ee
Now, $[\op X, \op p] = i \1$ implies $e^{-i\frac{\op H_{\omega_0}}{\beta}\op{p}}\op{X}e^{i\frac{\op H_{\omega_0}}{\beta}\op{p}} = \op{X} - \frac{\op H_{\omega_0}}{\beta}$ and, thus:
\be
    e^{-i\frac{\op H_{\omega_0}}{\beta}\op{p}} \abs{ f \lb(\op{X} + \frac{\op H_{\omega_0}}{\beta} \rb) }^2e^{i\frac{\op H_{\omega_0}}{\beta}\op{p}} = \abs{f(\op{X})}^2
\ee
Furthermore, $e^{-i\frac{\op H_{\omega_0}}{\beta}\op{p}}(\beta\op{X})e^{i\frac{\op H_{\omega_0}}{\beta}\op{p}} = \beta\op{X} - \op H_{\omega_0}$. Now, using two equivalent definitions of \emph{Connes cocycles} (see Appendix~C of \cite{Sorce}) we have $\op H_{\omega_h|\omega_0} - \op H_{\omega_0} = \op H_{\omega_h} - \op H_{\omega_0|\omega_h}$, where $\op H_{\omega_h}$ is the modular Hamiltonian corresponding to the state $|\omega_h\rangle$. Thus, \cref{comp2} simplifies to
\be \label{eq:intermediate-step}
    -\log{\op\rho_{\hat{\omega}_h}}  = e^{-i\frac{\op H_{\omega_0}}{\beta}\op{p}} (\op H_{\omega_h} - \op H_{\omega_0|\omega_h}) e^{i\frac{\op H_{\omega_0}}{\beta}\op{p}} + \beta\op{X} - \log\abs{ f(\op{X})}^2 + \log\beta 
\ee
Now note that separately $\op H_{\omega_h}, \op H_{\omega_0|\omega_h} \notin \mathfrak A(\hor_R, \omega_0)$. However $(\op H_{\omega_h} - \op H_{\omega_0|\omega_h}) \in \mathfrak A(\hor_R, \omega_0)$ and it follows from the fact that the operator is proportional to the derivative at identity of the Connes flow relating modular flows of $|\omega_0\rangle$ and $|\omega_h\rangle$. See Appendix~C of \cite{Sorce} for more details. Furthermore, for the state $|\omega_h\rangle$ in $\mathcal{P}^{\#}$ one has $j_{\omega_0}(\op U)\op\Delta_{\omega_0}j_{\omega_0}(\op U^*) = \op \Delta_{\omega_0|\omega_h}$ \cite{HI}. So, $\op H_{\omega_0|\omega_h} =j_{\omega_0}(\op U)\op H_{\omega_0}j_{\omega_0}(\op U^*)$ Now as $j_{\omega_0}(\op U) \in \mathfrak A(\hor_R, \omega_0)'$, so clearly the flows generated by $\op H_{\omega_0}$ and $\op H_{\omega_0|\omega_h}$ match on $\mathfrak A(\hor_R, \omega_0)$. Hence \cref{eq:intermediate-step} can be written as:
\be
    -\log{\op\rho_{\hat{\omega}_h}} = e^{-i\frac{\op H_{\omega_0|\omega_h}}{\beta}\op{p}} \op H_{\omega_h} e^{i\frac{\op H_{\omega_0|\omega_h}}{\beta}\op{p}} - \op H_{\omega_0|\omega_h} + \beta\op{X} - \log\abs{ f(\op{X})}^2 + \log\beta
\ee
Now recall that the von Neumann entropy of the classical-quantum state $\ket{\hat\omega_h}$ is:
\begin{equation} \label{von}
    S_{\text{vN}}(\op\rho_{\hat\omega_h}) = -\Tr (\op\rho_{\hat\omega_h}\log\op\rho_{\hat\omega_h}) = - \braket{ \hat\omega_h | \log\op\rho_{\hat\omega_h} |\hat\omega_h }
\end{equation}
Since the function $f$ is slowly varying in $X$, its Fourier transform is sharply peaked around momentum $p=0$ and hence $\op p|f\rangle \approx 0$. Using this and that $\op H_{\omega_h} \ket{\omega_h} = 0$, as every state has zero energy with respect to its modular Hamiltonian, we have in the slowly varying $f$ approximation:
\be
    \braket{ \hat{\omega}_h|e^{-i\frac{\op H_{\omega_0|\omega_h}}{\beta}\op{p}}\op H_{\omega_h}e^{i\frac{\op H_{\omega_0|\omega_h}}{\beta}\op{p}}|\hat{\omega}_h} \approx 0
\ee
Now:
\be
    \braket{ \hat{\omega}_h | (-\op H_{\omega_0|\omega_h}) |\hat{\omega}_h} = \braket{ \omega_h | \log\op \Delta_{\omega_0 |\omega_h} |\omega_h } = \braket{ \omega_h |-\log\op\Delta_{\omega_h | \omega_0} | \omega_h} = -S(\omega_h|\omega_0)
\ee
where the first equality is trivial and in the second equality we have used the fact that $|\omega_h\rangle$ is in $\mc{P}^{\#}$ and hence $\op J_{\omega_h}\op\Delta_{\omega_h|\omega_0}\op J_{\omega_h} = \op \Delta_{\omega_0|\omega_h}^{-1}$ (which follows because for $|\omega_h\rangle \in \mathcal P^{\#}$, the relative modular conjugation $\op J_{\omega_h|\omega_0} = \op J_{\omega_h}$ and the fact that $\op J_{\omega_h|\omega_0}\op\Delta_{\omega_h|\omega_0}\op J_{\omega_h|\omega_0} = \op \Delta_{\omega_0|\omega_h}^{-1}$. See sec.~V.2 of \cite{Haag} for more details). The final equality follows from the definition of relative entropy (\cref{eq:rel-entropy-defn}). Thus the von Neumann entropy (\cref{von}) evaluates to
\be\label{Masteratappendix}
    S_{\text{vN}}(\op\rho_{\hat{\omega}_h}) = -S(\omega_h|\omega_0) + \beta \braket{ \op X }_{\hat{\omega}_h} + S(f) + \log\beta
\ee
where $S(f) = -\int_{\mathbb{R}}dX|f(X)|^2\log|f(X)|^2$, as given in \cref{Master}.\\



\bibliographystyle{JHEP}
\bibliography{crossed-product}      
\end{document}